\documentclass[review, 3p, 11pt]{elsarticle}

\usepackage{graphicx}
\usepackage{color}
\usepackage{epsfig}
\usepackage{booktabs}
\usepackage{latexsym}
\usepackage{amsmath}
\usepackage{eufrak}
\usepackage{hyperref}
\usepackage{subfigure}
\usepackage{ulem}
\usepackage{lettrine}
\usepackage{verbatim}
\usepackage{array}
\usepackage{varwidth}
\usepackage{mathptmx} 
\usepackage{setspace}
\usepackage{natbib}

\biboptions{numbers, compress}

\journal{Microvascular Research}

\begin{document}

\begin{frontmatter}

\title{Vascular journey and adhesion mechanics of micro-sized carriers in narrow capillaries}

\author[potenza]{Alessandro Coclite\corref{cor}}
\ead{alessandro.coclite@unibas.it}

\cortext[cor]{Corresponding author}

\address[potenza]{Scuola di Ingegneria, Universit\`{a} degli Studi  della Basilicata, Viale dell'Ateneo Lucano -- 85100 Potenza, Italy}

\begin{abstract}

In this work a Lattice Boltzmann--Immersed Boundary method is used for predicting the dynamics of rigid and deformable adhesive micro-carriers (1 $\mu m$) navigating a capillary by the size of 10 $\mu m$ with 20\% hematocrit. Red cells and particles are modeled as a collection of mass-spring elements responding to a bending potential, an elastic potential and total enclosed area conservation constraint. Furthermore, particle surfaces are uniformly decorated with adhesive molecules (ligands) interacting with receptors disposed on the walls. \textcolor{blue}{Particle adhesion is modeled as a short-range ligad-receptor interaction and in term of formation and destruction probability functions that discriminate whether a chemical bond can be formed or destroyed. If a bond is established an attractive elastic force is activated.} Particle transport and adhesion are characterized in terms of their ability to reach the capillary peripheries (margination rate) and firmly adhere the vasculature. This analysis is carried out systematically by varying particles' and cells' releasing positions and stiffness (Ca = 0 and $10^{-2}$). Moreover, three rigid and soft representative particles are transported on a finer mesh ($\Delta x$ = 15 nm) and the chemical strength of their adhesive coating is varied ($\sigma$ = 0.5, 1.0, and 2.0) to precisely analyze the resulting adhesion mechanics. Stiffness is found to weakly influence the margination rate while significantly affect the ability of such constructs to efficiently interact with the endothelium by forming stable chemical bonds.

\end{abstract}


\end{frontmatter}

\section{\textsc{Introduction}}

Micro- and nano-particles have been proven as efficient carriers of therapeutics for the specific treatment of diseases such as cancer or cardiovascular disorders.~\cite{peer2007,antoniades2010} For the target specific delivery of drugs, two major steps are required: the accumulation of these small constructs into capillary peripheries (\textit{margination}) and the firm adhesion to the diseased tissue capillary walls. \textcolor{blue}{Nanomedicines should release their cargo mostly when firmly adhering capillary walls or being dragged into tumoral pore thorough a fenestration and then directly release their therapeutics cargo \textit{in situ} (\textit{extravasation}). Two routes have been proposed for particles extravasation, \textit{interendothelial} and \textit{transendothelial}.~\cite{podduturi2013,moghimi2018,vu2019} Indeed, understanding the specific mechanisms for controlling particle margination, adhesion and extravasation represent a major step for the rational design of anti-cancer nanopharmaceuticals.}

Bio-inspired computational methods are gaining increasing interest in the scientific community. The accurate and efficient numerical simulations of biological systems provide essential means in understanding the fundamental physics and reducing the time and cost needed for experiments. In this context, due to the large number of parameters involved in the vascular transport of platelets-like objects, computational methods are becoming of ever increasing interests. Particles can be precisely tailored in term of their shape, size, superficial properties and stiffness (the so called 4S parameters) in order to modulate their abilities.~\cite{decuzzi2010,decuzzi2016}    
In recent years, more and more of scientists have studied blood flows analyzing the complex features related to the presence of red blood cells, platelets and cells into the margination of micro- and nanoparticles.~\cite{muller2014,mountrakis2014,anselmo2015,spann2016} By definition, margination corresponds to the dislodging of immersed objects toward vessel walls. It has been observed experimentally for white cells, platelets and rigid microparticles, however the dependence of the margination rate on particles' stiffness, shape, size, and surface properties remains an open question. Then, such carriers should stably adhere to the vessel walls in order to support the continuous and controlled release of drugs into the diseased tissue.~\cite{coclite20181} The recipe for efficient micro- and nano-carriers is far from being discovered and computational models are useful tools for the design of such constructs. Indeed, reliable computational schemes must account for the transport of several structures with different stiffnesses, densities and shapes; as well as for the need of modeling particle-particle and particle-walls interactions.~\cite{coclite20183,coclite20191} 

In this work, an hybrid kinematic/dynamics Immersed--Boundary (IB)/Lattice Boltzmann (LB) scheme is employed.~\cite{coclite20191} \textcolor{blue}{In recent years, two approach emerged to enforce the boundary conditions on the immersed structure within IB schemes. The most classic kinematic~IB, in which no-slip boundary conditions are automatically imposed by dragging the Lagrangian markers with the withstanding fluid velocity and, in turn, the effect of the body presence on the surrounding fluid is taken into account by a forcing term added into the momentum equation.~\cite{peskin2002}. On the other hand, the dynamic~IB determines the Lagrangian markers velocity by solving the Newton law and enforces the no-slip conditions by introducing an effective forcing term accounting for the presence of the boundary. Indeed, the two approaches overlap for neutrally buoyant structures while only the dynamic IB is able to capture inertial bodies dynamics.~\cite{coclite20191}} Here, the immersed boundary technique is employed in its kinematic formulation for the red blood cells thus limitating the computational burden. Red cells are considered here as dense as plasma, so that, membranes velocity is advected from the withstanding fluid velocity. On the contrary, microparticles are transported in the 10 $\mu m$ narrow capillary with the dynamic IB formulation extensively developed and validated by the author and colleagues.~\cite{coclite20163,coclite20172} Such dynamic--IB scheme is adopted due to its ability to compute, within the same framework, rigid and deformable inertial particles with any buoyancy. Particles are chosen to be slightly denser than RBCs. \textcolor{blue}{Their dynamics is computed under different conditions, namely: four groups of red blood cells randomly distributed into the computational domain fulfilling the physiological hematocrit constraint; four particles per time are transported within each RBCs random initial configurations, each particle initial position is randomly determined and five groups of particles are considered; two values for particle's mechanical stiffness -- Ca = 0 and 10$^{-2}$; and three values for the particle chemical affinity with the vascular walls -- $\sigma$ = 0.5, 1.0, and 2.0.}

Firstly, blood cells are transported alone, thus identifying three flow regions discriminated by the perturbation induced into the flow field by their presence: a bulk region, RBC--rich zone; an intermediate layer, fluid region in which the flow is only slightly perturbed by the cells peripheries; and a cell--free layer, an essentially unperturbed flow region. Then, particles are released into the fluid domain and the distribution over time of the number of particles populating these three regions is computed. The margination ability of soft and rigid constructs is measured by counting the number of particles populating over time these three regions. Interestingly, for 1 $\mu m$ construct the stiffness poorly affects the margination rate while soft particles are more prone to firmly adhere the vasculature. As the matter of facts, only 6\% of the total number of rigid particles can firmly adhere to the vasculature while this number grows up to 23\% when considering soft membranes.

\section{\textsc{Computational Method}}

\subsection{\textsc{Two dimensional BGK--Lattice Boltzmann Method}}

The fluid is modeled in terms of a set of 9 discrete distribution functions, $[f_{i}],\ (i=0,\dots,8)$, which obey the discrete Boltzmann equation
\begin{equation}
{f_i(\vec{x}+\vec{e}_i\Delta t, t+\Delta t)-f_i(\vec{x}, t)=-\frac{\Delta t}{\tau}[f_i(\vec{x}, t)-f^{eq}_i(\vec{x}, t)]}\, ,
\label{BGK}
\end{equation}
in which $\vec{x}$ and $t$ are the spatial and time coordinates, respectively; $[\vec{e}_{i}],(i=0,...,8)$ is the vector of the 9 discrete velocities; $\Delta t$ is the time step; and $\tau$ is the relaxation time given by the unique non-null eigenvalue of the collision term in the BGK-approximation.~\cite{bgk} The kinematic viscosity is related to $\tau$ as $\nu=c_s^2\, (\tau-\frac{1}{2}) \Delta t$, being $c_s=\frac{1}{\sqrt{3}}\frac{\Delta x}{\Delta t}$ the reticular speed of sound. Macroscopic variables are obtained by the moments of the distribution functions: fluid density $\rho=\sum _{i}f_{i}$, velocity $\rho\vec{u}= \sum _{i}f_{i} \vec{e}_{i}$, and pressure $p=c_{s}^{2} \rho =c_{s}^{2} \sum _{i} f_{i}$. The local equilibrium density functions are expressed by the Maxwell-Boltzmann distribution projected on the lattice:
\begin{equation}
{f^{eq}_i(\vec{x},t)=\omega_i\rho \Bigl[ 1+\frac{1}{c_s^2}(\vec{e}_i\cdot \vec{u})+\frac{1}{2c_s^4}(\vec{e}_i\cdot \vec{u})^2-\frac{1}{2c_s^2}(\vec{u} \cdot \vec{u}) \Bigr] }\, .
\label{feq}
\end{equation}
The set of the nine discrete velocities is given by: 
\begin{equation}
{\vec{e}_i= \begin{cases}
(0,0)\, , & \quad if \quad i = 0 \\
\Biggl(\cos\Biggl(\frac{(i-1)\pi}{2}\Biggr),\sin\Biggl(\frac{(i-1)\pi}{2}\Biggr)\Biggr)\, , & \quad if \quad i = 1-4 \\
\sqrt{2}\Biggl(\cos\Biggl(\frac{(2i-9)\pi}{4}\Biggr),\sin\Biggl(\frac{(2i-9)\pi}{4}\Biggr)\Biggr)\, , & \quad if \quad i = 5-8 \\
\end{cases}}
\label{GaussHermite}
\end{equation}
with the weight, $\omega_i=1/9$ for $i=1-4$, $\omega_i= 1/36$ for $i=5-8$, and $\omega_0=4/9$. 
The boundaries of the computational domain are treated with the Zou and He known velocity bounce back conditions.~\cite{zouhe1997}

\subsection{\textsc{Immersed Boundary Treatment}}

The immersed body consists in a network of $nv$ vertices linked with $nl$ linear elements, whose centroids are usually referred as \textit{Lagrangian markers}. Following the forcing procedure by Guo et al~\cite{guo2002} a forcing term $[\mathcal{F}_{i}] (i=0,...,8)$ is included as an additional contribution on the right-hand side of Eq.\eqref{BGK}:
\begin{equation}
{f_i(\vec{x}+\vec{e}_i\Delta t, t+\Delta t)-f_i(\vec{x}, t)=-\frac{\Delta t}{\tau}[f_i(\vec{x}, t)-f^{eq}_i(\vec{x}, t)]+\Delta t \mathcal{F}_{i}}\, .
\label{forcedBGK}
\end{equation}
$\mathcal{F}_{i}$ is expanded in term of the reticular Mach number, $\frac{\vec{e}_{i}}{c_{s}}$, resulting in:
\begin{equation}
{\mathcal{F}_{i}=\Biggl(1-\frac{1}{2\tau}\omega_i \Bigl[\frac{\vec{e}_i-\vec{u}}{c_s^2}+\frac{\vec{e}_i\cdot \vec{u}}{c_s^4}\vec{e}_i \Bigr] \Biggr)  \cdot \vec{f}_{ib}}\, ,
\label{ForcingTerm}
\end{equation}
where $\vec{f}_{ib}$ is a body force term. Due to the presence of the forcing term, the momentum density is derived as $\rho\vec{u}= \sum _{i}f_{i} \vec{e}_{i}+\frac{\Delta t}{2}\vec{f}_{lb}$.

In the present work, $\mathcal{F}_{i}$ accounts for the presence of an arbitrary shaped body immersed into the flow field. 
The Immersed Boundary (IB) procedure, extensively proposed and validated by Coclite and collaborators \cite{coclite20191}, is here adopted and the moving-least squares reconstruction by Vanella et al. \cite{vanella2009} is employed to exchange all Lattice Boltzmann distribution functions between the Eulerian lattice and the Lagrangian chain, while the body force term in Eq.\eqref{ForcingTerm}, $\vec{f}_{ib}$, is evaluated through the formulation by Favier et al.~\cite{pinelli2014}.

\textbf{\textit{Elastic Membrane Deformation.}} Membranes are subject to elastic strain response, bending resistance, and total enclosed area conservation. The stretching elastic potential acting on the two vertices sharing the $l$-th element is given as
\begin{equation}
{V_{l}^{s}=\frac{1}{2}k_{s}(l_{l}-l_{l,0})^{2}}\, ,
\label{strainPot}
\end{equation}
being $k_{s}$ the elastic constant, $l_{l}$ the current length of the $l$-th element, and $l_{l,0}$ the length of the $l$-th element in the stress-free configuration. The nodal forces corresponding to the elastic energy for nodes 1 and 2 connected by \textit{l} reads:
\begin{equation}
{\begin{cases}
\vec{F}_{1}^{s}=-k_{s}(l-l_{0})\frac{\vec{r}_{1,2}}{l}\, , \\
\vec{F}_{2}^{s}=-k_{s}(l-l_{0})\frac{\vec{r}_{2,1}}{l}\, , \\
\end{cases}}
\label{strainFor}
\end{equation}
where $\vec{r}_{i,j}=\vec{r}_{i}-\vec{r}_{j}$ and $r_{i}$ is the position vector of \textit{i} with respect to \textit{j}.

The bending resistance related to the $v$-th vertex connecting two adjacent element is
\begin{equation}
{V_{v}^{b}=\frac{1}{2}k_{b}(k_{v}-k_{v,0})^{2}}\, ,
\label{bendPot}
\end{equation}
with $k_{b}$ the bending constant, $k_{v}$ the current local curvature; $k_{v,0}$ the local curvature of the stress-free configuration. The local curvature is computed by measuring the variation of the angle between the two adjacent elements ($\theta -\theta _{0}$), with $\theta _{0}$ the angle in the stress free configuration. The resulting forces on the nodes $v_{left}$, $v$, and $v_{right}$ are:
\begin{equation}
{\begin{cases}
\vec{F}_{v_{left}}^{b}=k_{b}(\theta -\theta_{0})\frac{l_{left}}{l_{left}+l_{right}}\vec{n}_{v}\, , \\
\vec{F}_{v}^{b}=-k_{b}(\theta -\theta _{0})\vec{n}_{v}\, , \\
\vec{F}_{v_{right}}^{b}=k_{b}(\theta -\theta_{0})\frac{l_{right}}{l_{left}+l_{right}}\vec{n}_{v}\, , \\
\end{cases}}
\label{bendFor}
\end{equation}
where $l_{right}$ and $l_{left}$ are the length of the two adjacent left and right edges, respectively, and $\vec{n}_{v}$ is the outward unity vector centered in \textit{v}. In this model for the bending resistance the relation between the strain response constant $k_{s}$ and $k_{b}$ is expressed through $E_{b}=\frac{k_{b}}{k_{s}r^{2}}$, with $r$ the particle radius.

In order to constrain the enclosed area a penalty force is expressed in term of the reference pressure $p_{ref}$ and directed along the normal inward unity vector of the $l$-th element $(\vec{n}_{l})^{-}$: 
\begin{equation}
{\vec{F}_{l}^{a}=-k_{a}(1-\frac{A}{A_{0}})p_{ref} (\vec{n}_{l})^{-} l_{l}}\, ,
\label{areaFor}
\end{equation}
with $l_{l}$ the length of the selected element, $k_{a}$ the incompressibility coefficient, $A$ the current enclosed area, $A_{0}$ the enclosed area in the stress-free configuration. The enclosed area is computed using the Green's theorem along the curve $A=\frac{1}{2}\sum_l x_l dy_l$. Within this formulation $k_{a}=1$ returns a perfectly incompressible membrane. Note that $\vec{F}_{l}^{a}$ is evenly distributed to the two vertices connecting the \textit{l}-th element ($v_{left}$ and $v_{right}$) as $\vec{F}_{l}^{a}=0.5\vec{F}_{v_{left}}^{a}+0.5\vec{F}_{v_{right}}^{a}$.

\textbf{\textit{Particle-Particle Interaction}} Two-body interactions are modeled through a purely repulsive potential centered in each vertex composing the immersed particles. The repulsive force is such that the minimum allowed distance between two vertices coming from two different particles is $\Delta x$. The impulse acting on vertex $1$, at a distance $d_{1,2}$ from the vertex $2$ of an adjacent particle is directed in the inward normal direction identified by $(\vec{n}_{1})^{-}$ and is given by:
\begin{equation}
{\vec{F}_{1}^{pp}=\frac{10^{-4}}{8\sqrt{2}}\sqrt{\frac{\Delta x}{d_{1,2}^{5}}}(\vec{n}_{1})^{-}}\, .
\label{repFor}
\end{equation}

\textbf{\textit{Wall-Particle Interaction}} With no intent of being exhaustive, a wide range of vascular molecules are involved in the adhesion process, including E- and P-selectins, $\alpha_v \beta_3$ and $\alpha_v \beta_5$ integrins, VCAM-1 and ICAM-1 adhesion molecules.~\cite{burdick2003,barthel2007,myung2011} In the present work, the author only consider the effect of ICAM-like adhesion molecules, meaning a shot-range ligand-receptor interaction. Ligand molecules are distributed over particle boundary with density $\rho_l$ and receptors are uniformly distributed over walls.~\cite{coclite20172} Ligand molecules are modeled as linear springs establishing bonds with receptors, the resulting force is given as
\begin{equation}
{\vec{F}_{l}^{wp}=\sigma (y_{l}-y_{cr,eq})\vec{n}_{l}}\, ,
\label{adhFor}
\end{equation}
where $y_{l}$ is the bond length, $y_{cr,eq}$ the equilibrium bond length and $\sigma $ the spring constant (same for all springs). Bonds can be generated only if the minimum separation distance between the particle boundary and the wall is smaller than a critical value, $y_{cr}=25\, nm$~\cite{roy2010}. The equilibrium bond length is $y_{cr,eq}$= 0.5$y_{cr}$ and gives a null force. The spring constant, $\sigma$ is computed in \textit{lattice units} through its dimensionless group, $\frac{\rho _{ref}\nu_{ref}^{2}}{H}$, where $\rho _{ref}$, $H$, and $\nu_{ref}$ are the reference density, length, and kinematic viscosity, respectively. At each time step, the bond formation is regulated by a forward probability function, while a reverse probability function controls the destruction of a pre-existing bond, see for details \cite{coclite20172}. The adhesive force, being calculated at the centroid of each element, is evenly distributed to the two vertices connecting the element in the same fashion used for $\vec{F}^{a}$.

\textbf{\textit{Hydrodynamics Stresses}} Pressure and viscous stresses exerted by the $l$-th linear element are:
\begin{equation}
{\vec{F}_{l}^{p}(t)=(-p_{l} \vec{n}_{l})l_{l}}\, , \\
\label{hydroForP}
\end{equation}
\begin{equation}
{\vec{F}_{l}^{\tau }(t)=(\bar{\tau }_{l}\cdot \vec{n}_{l})l_{l}}\, ,
\label{hydroForNu}
\end{equation}
where $\bar{\tau }_{l}$ and $p_{l}$ are the viscous stress tensor and the pressure evaluated in the centroid of the $l$-th element, respectively. The pressure and velocity derivatives in Eq.s \eqref{hydroForP} and \eqref{hydroForNu} are evaluated using a probe in the normal positive direction of each element, being the probe length $1.2\, \Delta x$, and using the cited moving least squares formulation \cite{vanella2009}. In this framework, the velocity derivatives evaluated at the probe are considered equal to the ones on the linear element centroid and all force contributions are computed with respect to the centroid of each elements and then transferred to the vertices.~\cite{MDdTJCP2016}

\subsection{\textsc{Fluid-Structure interaction}}

The motion of the immersed bodies is described differently for red cells and particles, for saving computational costs. On one side, blood cells are transported using the \textit{kinematic IB} technique described in \cite{coclite20191}; so that, advecting the velocity of the Lagrangian points with the underlying Eulerian fluid velocity and then applying internal stresses (Eq.s \eqref{strainFor}, \eqref{bendFor}, \eqref{areaFor}, and \eqref{repFor}) through the volume force in the flow equations using Eq.\eqref{ForcingTerm}. On the other side, the particles dynamics is determined by \textit{dynamics IB} technique described in \cite{coclite20191}, so that using the solution of the Newton equation for each Lagrangian vertex, accounting for both internal, Eq.s \eqref{strainFor}, \eqref{bendFor}, \eqref{areaFor}, \eqref{repFor}, and \eqref{adhFor} and external stresses, Eq.s \eqref{hydroForP} and \eqref{hydroForNu}. Then, no-slip boundary conditions are imposed using a weak coupling approach. \cite{coclite20163}

\textbf{\textit{Blood Cells Motion.}} For each vertex nine Eulerian points are considered, namely the Eulerian points falling into a square with side equal $2.6 \Delta x$ and the velocity of each Lagrangian point, $\dot{\vec{x}}(t)$, is computed interpolating the velocity of the fluid from the nine associated Eulerian points. Finally, the position of each Lagrangian point is computed as:
\begin{equation}
{\vec{x}(t)=\frac{2}{3}(2\vec{x}(t-\Delta t)-\frac{1}{2}\vec{x}(t-2\Delta t)+\dot{\vec{x}}(t)\Delta t)}\, .
\label{kinematicPos}
\end{equation}

\textbf{\textit{Elastic Particles Motion.}} The total force $\vec{F}_{v}^{tot}(t)$ acting on the $v$-th element of the immersed body is evaluated in time and the position of the vertices is updated at each Newtonian dynamics time step considering the membrane mass uniformly 
distributed over the $nv$ vertices,
\begin{equation}
{m_{v}\dot{\vec{u}}_{v}=\vec{F}_{v}^{tot}(t)=\vec{F}_{v}^{s}(t)+\vec{F}_{v}^{b}(t)+\vec{F}_{v}^{a}(t)+\vec{F}_{v}^{pp}(t)+\vec{F}_{v}^{wp}(t)+\vec{F}_{v}^{p}(t)+\vec{F}_{v}^{\tau}(t)}\, .
\label{newtonSoft}
\end{equation}

The Newton equation of motion is integrated with the Verlet algorithm using, as first tentative velocity, the velocity obtained by interpolating the fluid velocity from the surrounding lattice nodes, $\dot{\vec{x}}_{v,0}(t)$, 
\begin{equation}
{\vec{x}_{v}(t+\Delta t)=\vec{x}_{v}(t)+\dot{\vec{x}}_{v,0}(t)\Delta t+\frac{1}{2}\frac{\vec{F}_{v}^{tot}(t)}{m_{v}}\Delta t^{2}+O(\Delta t^{3})}\, ,
\label{verletPos}
\end{equation}
then, the velocity at the time level $t+\Delta t$ is computed as
\begin{equation}
{\vec{u}_{v}(t+\Delta t)=\frac{\frac{3}{2}\vec{x}_{v}(t+\Delta t)-2\vec{x}_{v}(t)+\frac{1}{2}\vec{x}_{v}(t-\Delta t)}{\Delta t}+O(\Delta t^{2})}\, .
\label{verletVel}
\end{equation}

\textbf{\textit{Rigid Particles Motion.}} Rigid motion is readily obtained integrating all stresses contributions over the particles boundary and updating both, linear and angular velocity in time as $\dot{\vec{u}}(t)=\frac{\vec{F}^{tot}(t)}{m}$ and $\dot{\omega}(t)=\frac{M^{tot}(t)}{I}$. Here $m$ is the particle mass, $\vec{F}^{tot}(t)$ is the total force exerted by the particle; $M^{tot}(t)$ is the total moment acting on the particle, and $I$ is the moment of inertia. Finally, $\vec{u}(t)$ and $\omega (t)$ are computed as
\begin{equation}
{\vec{u}(t)=\frac{2}{3}(2\vec{u}(t-\Delta t)-\frac{1}{2}\vec{u}(t-2\Delta t)+\dot{\vec{u}}(t)\Delta t)+O(\Delta t^{2})}\, ,
\label{linearVelRig}
\end{equation}
\begin{equation}
{\omega (t)=\frac{2}{3}(2\omega (t-\Delta t)-\frac{1}{2}\omega 
(t-2\Delta t)+\dot{\omega }(t)\Delta t)+O(\Delta t^{2})}\, .
\label{angularVelRig}
\end{equation} 

\section{\textsc{Set--Up and boundary conditions.}} 

\begin{figure}[h]
\centering
\includegraphics[scale=0.25]{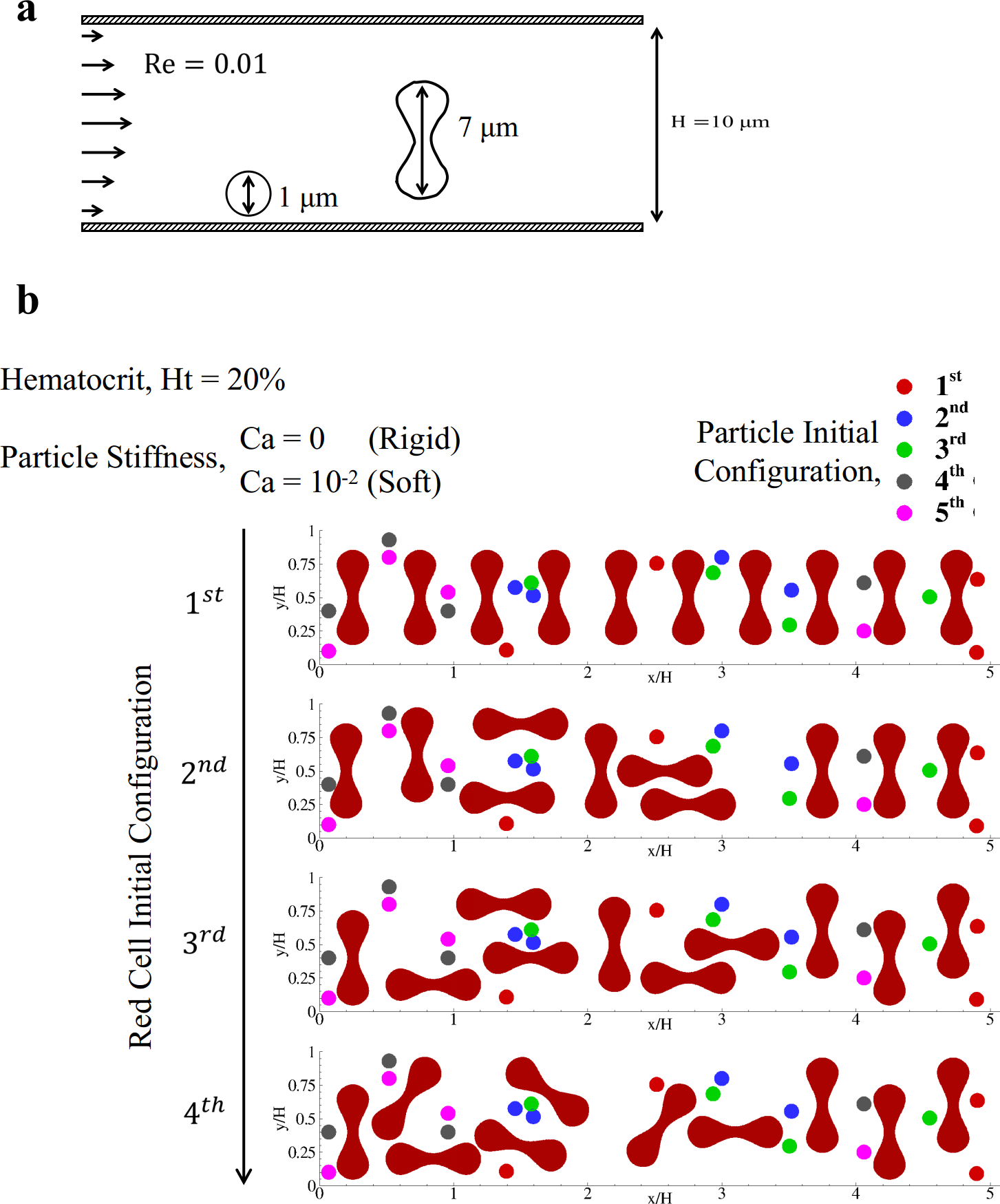}
\caption{{\bf Schematic of the physical problem.} {\bf a.} Sketch of the computational domain with characteristic length and dimensions. {\bf b.} The initial conditions for particles and red cells are obtained by choosing one per time the five particle initial positions within each of the four red blood cell configurations.}
\label{Schematic}
\end{figure}

The dynamics of particles and red cells navigating a narrow capillary ({\bf Figure.\ref{Schematic}}) is computed within a 2D rectangular computational domain by the height H $= 10 \mu m$ and length 5 H. Periodic boundary conditions are imposed along $x$ while no-slip conditions are adopted for the walls along y ({\bf Figure.\ref{Schematic}}a). A plane Hagen-Poiseuille flow is established by imposing a linear pressure drop $\Delta p$ ($=\frac{8u_{\max}^{2}}{H}\frac{\rho x}{Re}$) along the channel as function of the Reynolds number $Re=\frac{u_{\max }H}{\nu}=10^{-2}$. The reference kinematic viscosity is chosen as the water kinematic viscosity, $\nu $=1.2$\times $10$^{-6}$ m$^{2}$/s, thus resulting in the centerline velocity $u_{\max }=1.2\times 10^{-3} m/s$. The channel height is discretized with $200$ computational cells, corresponding to $\Delta x = 50 nm$. The relaxation time ($\tau=\frac{\nu}{c_s^2} + \frac{1}{2}$) related to the kinematic viscosity is fixed equal to unity for all of the present computations. Five initial configurations composed by four particles each are placed initially at the rest into the computational domain along with four different distributions of red cells ({\bf Figure.\ref{Schematic}}.b). The portion of fluid enclosed into RBC boundaries is about the 20\% of the total computational domain. Both, red blood cells and particles are discretized with a number of linear elements to obtain the ratio between solid Lagrangian and fluid Eulerian meshes as $0.3 \Delta x$. The resting configuration of the red blood cells is parameterized as: 
\begin{equation}
{\begin{cases} 
x=a\alpha \sin q \\
y=a\frac{\alpha }{2}(0.207+2.003 sin^{2}q-1.123 sin^{4}q)\cos q \end{cases}}\,
\label{rbcShape}
\end{equation}
where $a$ is a constant equal to 1.0, $\alpha$ is the cell radius, $3.5\, \mu m$, and $q$ varies in $\lbrack -0.5\pi , 1.5\pi \rbrack $\cite{shi2012}.  

Red cells' stiffness is regulated by the capillary number, Ca$_{rbc}$=$10^{-2}$ ($=\rho \nu u_{\max }/k_{s}$) and null bending resistance, $Eb_{rbc}=0$. The membrane stiffness is assumed as $k_s = 1.5\times 10^{-4}$ N/m. Note that, by assuming Ca$_{rbc}$=$10^{-2}$ within the above definition corresponds to assume Ca$_{rbc}$=0.92 with the more typical definition: $Ca=\rho \nu \dot{\gamma}\, a/E_{s}$ in which $\dot{\gamma}$ is the wall shear rate, $a$ is the RBC radius ($a=3.5 \mu m$), and $E_{s}$ is the shear modulus of the RBC membrane $E_s = 5.5\times 10^{-6}$ N/m~\cite{pozrikidis2005,omori2012,tomaiuolo2014}. 

On the other side, particles are circular membranes with Ca$_{p}$=$10^{-2}$ and $Eb=0.05$ ($=k_{b}/k_{s}r^{2}$), where $r$ ($=0.5 \mu m$) is the particles radius; or transported as rigid objects. This choice led the analysis of the interaction between small objects (particles) with large cells (RBCs) with about the same mechanical stiffness. Then, particle surfaces are equipped by ligand molecules with density $\rho_{L}$ = 0.5, meaning that the half of the particles' boundary is covered by ligands. These ligands interact with receptors uniformly distribuited on the capillary walls forming ligand-receptor bonds with an affinity given by $k_{f}/k_{r,0 }=8.5 \times 10^{3}$ and a bond strength of $\sigma =\frac{\varrho _{ref}\nu _{ref}^{2}}{H}=$ 1.0.~\cite{sun2008} This values return a plausible agreement with experimental observations and was validated against experimental data in two previous work by the author and collaborators \cite{coclite20172,coclite20181}. The solid density of particle membranes is considered as slightly higher than that of the surrounding fluid, $\frac{\rho _{P}}{\rho} =1.1$. Finally, thermal fluctuations are neglected being almost uninfluential for micrometric and sub-micrometric particles under flow.

\section{\textsc{Results and Discussion}}

\subsection{\textsc{Flow regimens in 10 $\mu m$ capillary}} 

\begin{figure}[h]
\centering
\includegraphics[scale=0.3]{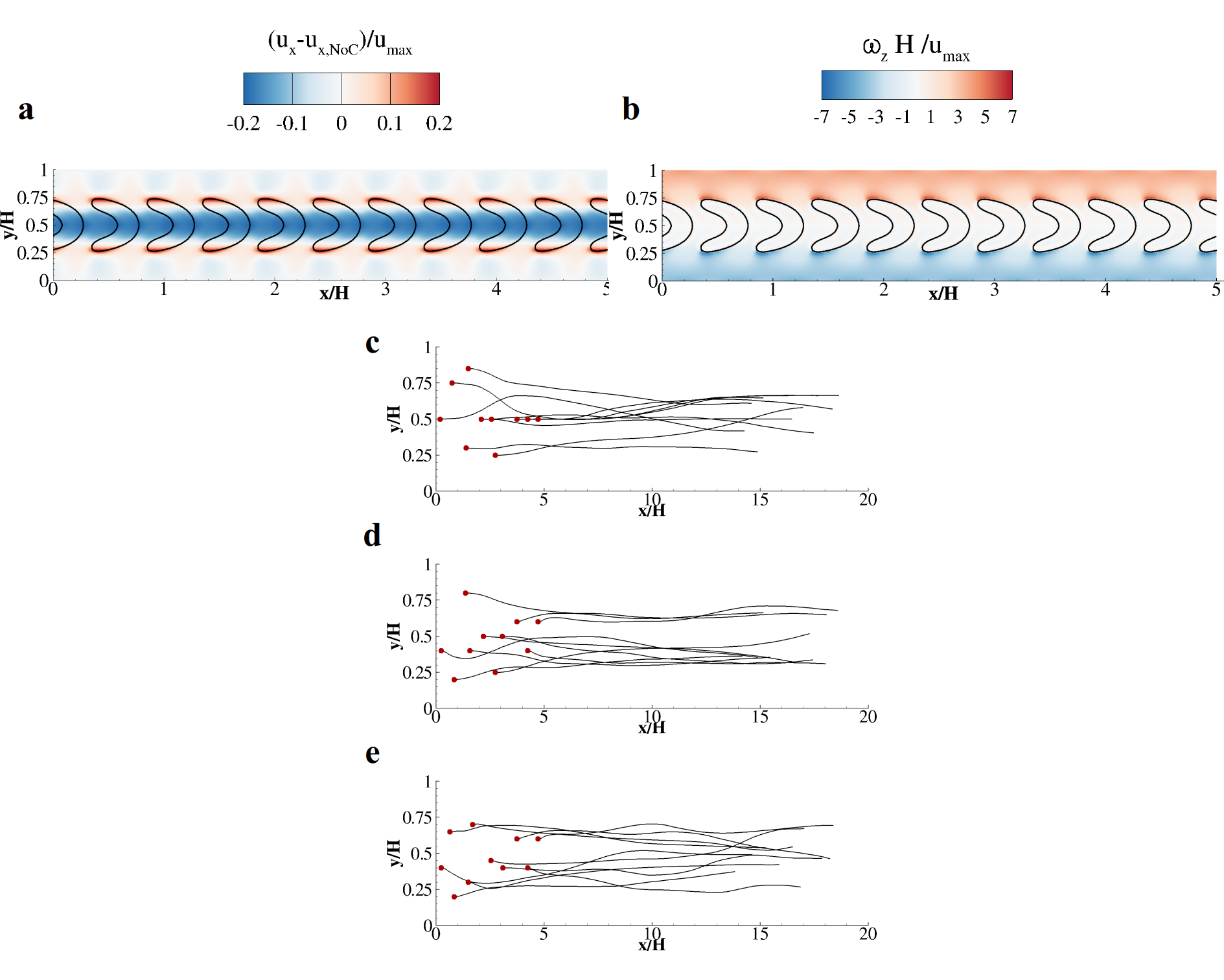}
\caption{{\bf Cell free layer formation in a 10 $\mu m$ high capillary with 20\% hematocrit.} {\bf a.} Contour plot of the linear slip velocity field along $x$, $\frac{u_{x}-u_{x,NoC}}{u_{{max}}}$, taken at $\frac{t\, u_{max}}{H}=$ 10, obtained with the initial ``single-filed'' RBC distribution. {\bf b.} Contour plot of the out of plane vorticity, $\frac{\omega_{z} H}{u_{max}}$, taken at $\frac{t\, u_{max}}{H}=$ 10, obtained with ``single-filed'' RBC distribution. ({\bf c}, {\bf d}, {\bf e}) Trajectories of the center of mass obtained with the three disordered initial distribution of RBCs for $0<\frac{t\, u_{max}}{H}<10$.}
\label{CFL}
\end{figure} 

Due to the great separation in the occupied fluid area between red blood cells and particles ($A_{RBC,Tot}/A_{part,Tot}$ = 34.74), particle-particle collisions are rare events while RBC-particle interactions dominate the redistribution of particles into the flow field. As a consequence, this RBC redistribution governs the journey of particles into the blood stream.~\cite{fitzgibbon2015,fedosov2014} For this reason, the transport of red blood cells alone is firstly considered and analyzed in terms of the overall cell configurations and trajectories (\textbf{Figure.\ref{CFL}}). The simplest initial configuration possible, depicted in \textbf{Figure.\ref{Schematic}.b}, is a ``single-filed'' distribution of RBCs. Indeed, the intrinsically equilibrated flow field led the cells to null lateral displacement while reaching the typical parachute shape (\textbf{Figure.\ref{CFL}.a}).~\cite{ma2009} 
Here, the presence of immersed cells perturb the flow by forming different layers, as demonstrated in \textbf{Figure.\ref{CFL}.a}. The slip velocity field, computed as the difference between the velocity field obtained when transporting the cells ($\frac{u_x}{u_{max}}$) and the velocity field with no cells immersed in ($\frac{u_{x,NoC}}{u_{max}}$), clearly shows the effect of the presence of elastic membranes into the flow. Precisely, the membrane responses to the exerted hydrodynamics forces is wide and negative into the higher velocity fluid laminae, localized and positive at the cells peripheries (see \textbf{Figure.\ref{CFL}.a}). The formation of such layers is strongly emphasized by seeing at the out of plane vorticity patterns (reported in \textbf{Figure.\ref{CFL}.b}). Here, the higher velocity fluid laminae are characterized by almost null vorticity values; on the contrary, cell extrema present the highest values. One can identify three fluid regions: a bulk zone, RBC-rich zone (4 $\mu m$) in which the flow is strongly affected by the presence of large cells immersed in, $0.3\le y/H\le 0.7$; an intermediate zone, 1.0 $\mu m$ high layer, in which the flow is slightly perturbed by the presence of cells peripheries, $0.2<y/H<0.3$ and $0.7<y/H<0.8$; and the cell-free layer, region of essentially undisturbed flow, $0<y/H<0.2$ and $0.8<y/H<1$. Typically, the cell-free layer and the intermediate zone are considered as a single fluid lamina by the height of 3 $\mu m$ for a capillary flow at Re = 0.01 in a 10 $\mu m$ vessel with 20\% hematocrit while for the scope of the present work the two flow regions are considered separately.~\cite{zhang2009,fedosov2010} 
  
With the idea of shuffling the RBCs distribution into the flow field, three different disordered initial configurations are computed in addition, see \textbf{Figure.\ref{Schematic}.b}. Large cells tend to occupy higher velocity regions regardless to their releasing lateral position as demonstrated by observing the cells trajectories (\textbf{Figure.\ref{CFL}.c}, \textbf{\ref{CFL}.d}, and \textbf{\ref{CFL}.e}). Immersed membranes tend to equilibrate the velocity and pressure fields across their boundary; being the plane Hagen-Poiseuille flow symmetric with respect to the centerline, large cells tend to move across flow streamlines in order to have their upper and lower hemi-surfaces equally distributed below and above the centerline. This phenomenon is strongly emphasized by the ratio between biconcave membrane size and the channel height (the blockage ratio) due to the large difference in the dragging fluid velocity across their boundaries.~\cite{shi2012} 

\subsection{\textsc{Microconstructs margination as a function of red blood cells distribution.}} 

\begin{figure}[h]
\centering
\includegraphics[scale=0.25]{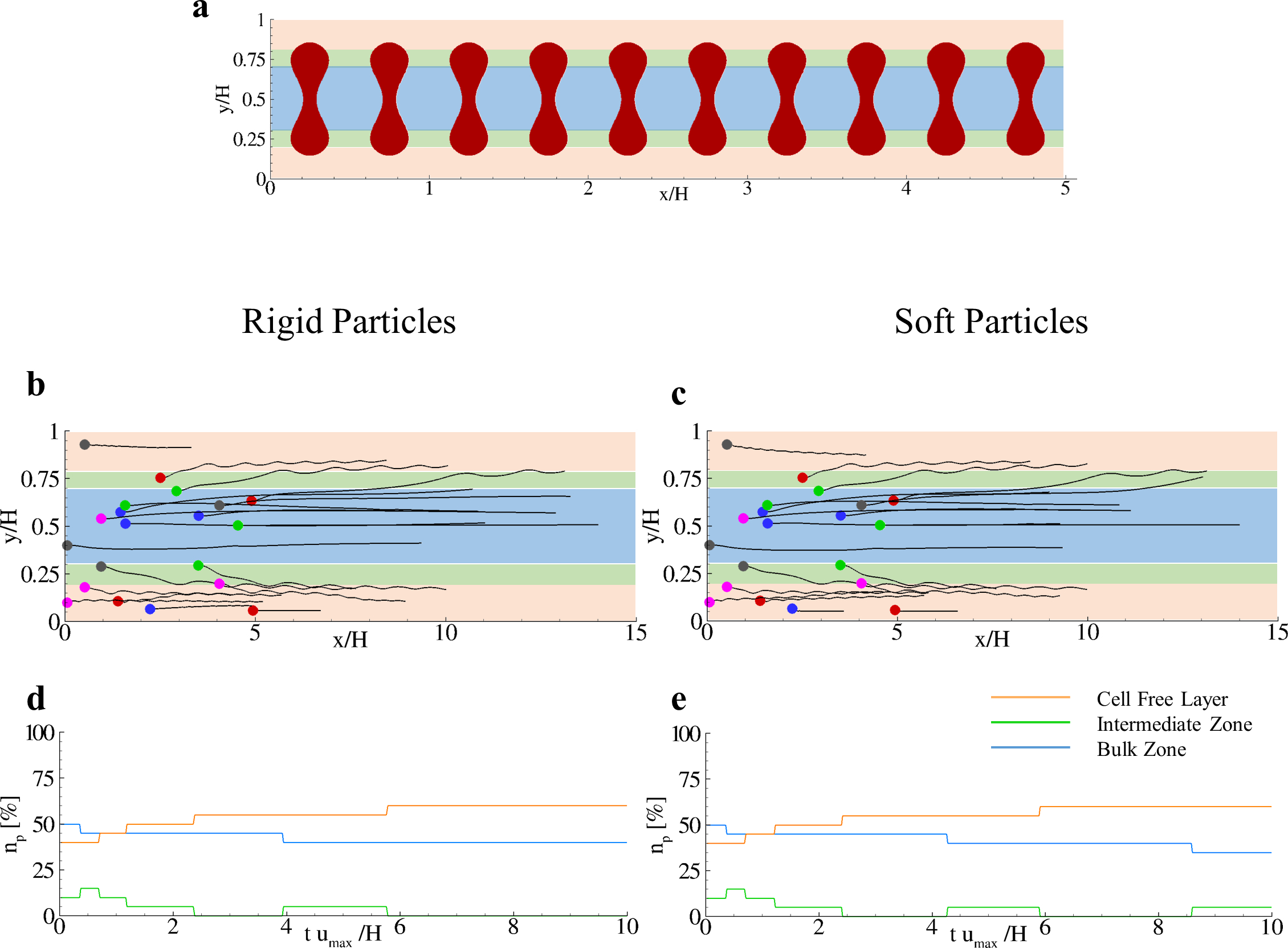}
\caption{{\bf Particle margination from a ``single-filed'' red blood cell initial distribution.} {\bf a.} Red cell releasing positions in the computational domain. ({\bf b}, {\bf c}) Particle trajectories when transported as rigid membranes (Ca = 0) and soft membranes (Ca = $10^{-2}$). ({\bf d}, {\bf e}) Number of particles populating the bulk zone, the intermediate zone and the cell-free layer over time.}
\label{singleFiled}
\end{figure}

The vascular journey of rigid and soft particles is now considered in terms of their trajectories and measuring the number of particles populating the three identified fluid regions. Firstly, the five different configurations of rigid (Ca = 0) and soft (Ca = $10^{-2}$) particles are transported one per time within the ``single-filed'' RBC distribution, as depicted in \textbf{Figure.\ref{Schematic}.b}. All particles (regardless from their rigidity) tend to avoid direct bumps with RBCs and accommodate themselves outside the perturbed flow regions; specifically, or between two RBCs or outside the bulk zone  (see \textbf{Figure.\ref{singleFiled}.b} and \textbf{Figure.\ref{singleFiled}.c} for Ca = 0 and Ca = $10^{-2}$, respectively). Quantitatively, about 50\% of the particles are initially released into the bulk zone along with 40\% into the cell-free layer and 10\% in the intermediate zone; then, after $6\, tu_{\max }/H$ the number of rigid particles in the margination layer increases by 20\%, the intermediate zone becomes completely unpopulated and the bulk loses about 10\% of particles (\textbf{Figure.\ref{singleFiled}.d}). Analogous trends are found when transporting the five different particle configurations with Ca = $10^{-2}$ (\textbf{Figure.\ref{singleFiled}.e}). \textcolor{red}{Indeed, for the ``single-filed'' RBCs case, particles' mechanical stiffness do not play any major role into their margination ability. In particular, red cells navigate the bulk region in an ordered disposition and being the particles' size largely smaller than the cells', particle-cell collisions could be considered as rare events.}

\begin{figure}
\centering
\includegraphics[scale=0.3]{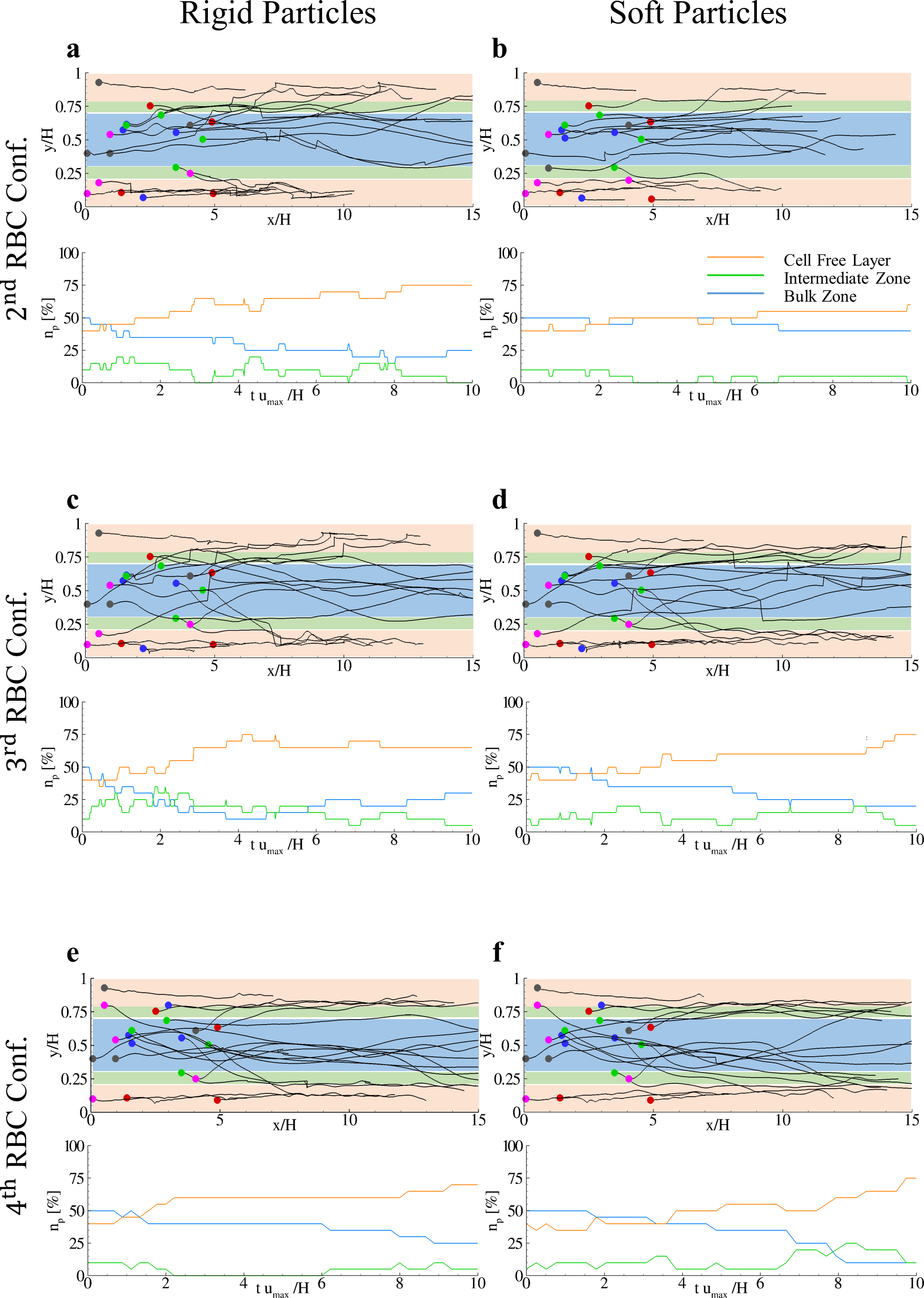}
\caption{{\bf Particle margination in three different red blood cell initial distributions.} Particle center of mass trajectories and number of particles in the bulk zone, the intermediate zone and the cell-free layer for rigid (Ca = 0) and soft particles (Ca = $10^{-2}$) obtained with the second ({\bf a}, {\bf b}), third ({\bf c}, {\bf d}) and fourth ({\bf e}, {\bf f}) initial disordered configurations of red blood cells (see {\bf Figure.\ref{Schematic}}).}
\label{randomInitial}
\end{figure}

On the contrary, when the particles are transported into disordered RBC distributions, the interactions between particles and cells become more relevant and slight changes in the RBC releasing positions strongly affect particle trajectories. Rigid particles trajectories present discontinuities due to elastic bumps with RBCs because of their impossibility to adsorb the exchanged momentum (top insets in \textbf{Figure.\ref{randomInitial}.a}, {\bf .\ref{randomInitial}.c}, and {\bf .\ref{randomInitial}.e}); on the contrary, the elastic capabilities of soft particles give smoother trajectories and the number of contacts with cells (discontinuities in the trajectories) is dramatically limited (top insets in \textbf{Figure.\ref{randomInitial}.b}, {\bf .\ref{randomInitial}.d}, and {\bf .\ref{randomInitial}.f}). Interestingly, switching to a disordered initial RBC configuration poorly affects the measured final number of particles populating the three identified flow regions with respect to the ``single-filed'' RBC positioning regardless of their stiffness. Nonetheless, disordered red blood cell initial conditions return different functional behavior for the particles distributions due to the increased number of RBC-particle interactions. Indeed, straightforward cell dynamics led to almost monotonic variations of the number of particle in the three regions while complex cell dynamics led to non-smooth distributions. Note that, the idealized configurations of the computational domain (straight rectangular channel with periodic boundary conditions) neglect the recombination effect of the RBC distributions caused by enlargements or restrictions into blood vessels. This recombination effect may play a major role in the margination mechanics, continuously shuffling the red blood cell distributions in the vasculature. \textcolor{red}{In this context, the geometrical complexity of physiological vasculatures would generate the RBCs recombination and significantly enhance the differences between rigid and soft particles in terms of their margination ability. As already demonstrated with experimental and computational findings by several researchers, the margination-adhesion mechanism strongly depends on the presence of large cells navigating the considered vasculature. Specifically, the interaction between cells and particles causes the particles being dislodged away from the bulk region.~\cite{anselmo2015,spann2016,melchionna2011} Indeed, neglecting the contribution of whole blood in the analysis of particles' dynamics can significantly underestimate their margination and vascular deposition.}

\subsection{\textsc{Particle regimens in 10 $\mu m$ high blood capillary.}} 

\begin{figure}
\centering
\includegraphics[scale=0.32]{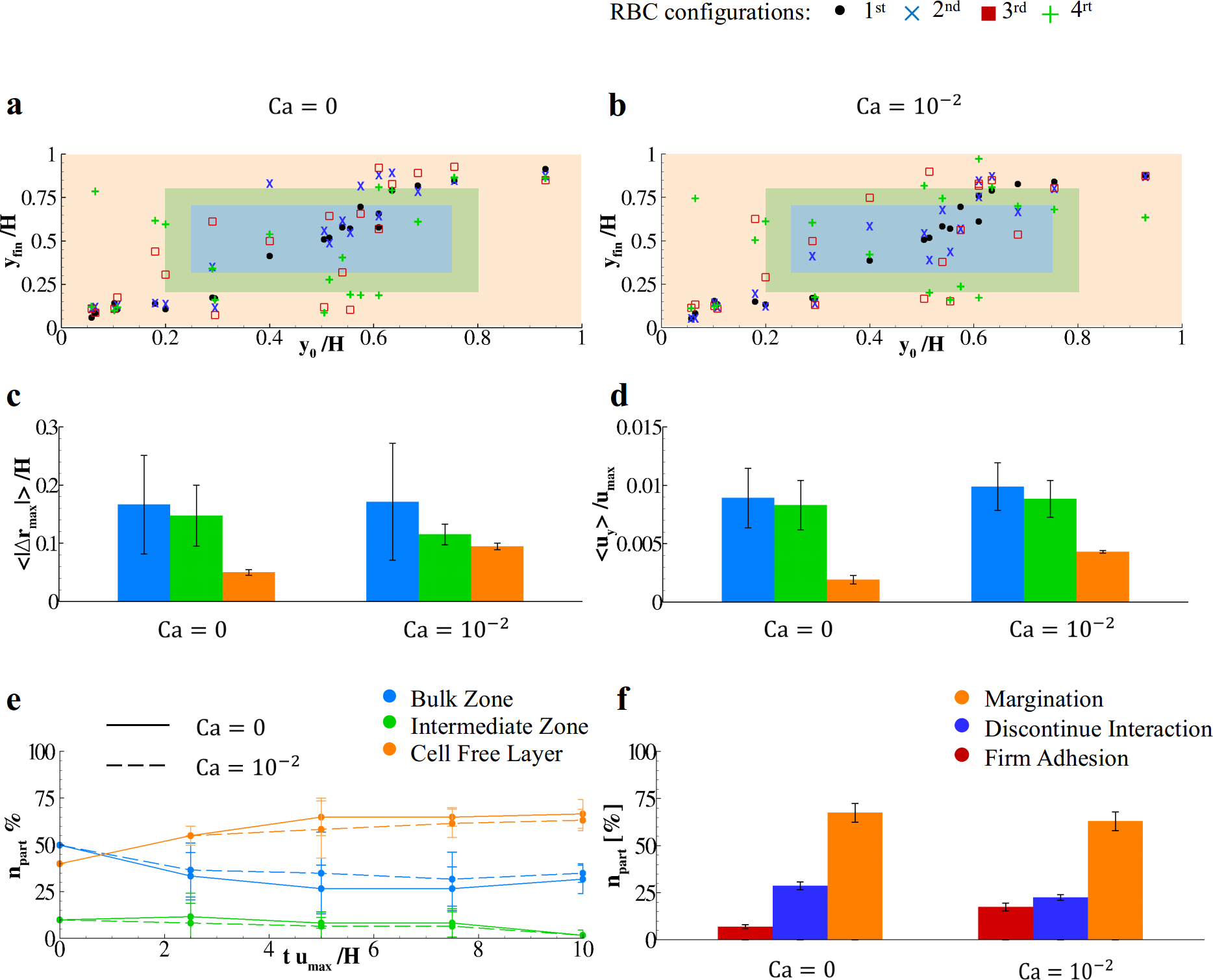}
\caption{{\bf Characterization of particle regimens in 10 $\mu m$ high blood capillaries.} ({\bf a}, {\bf b}) Plot of rigid (left) and soft (right) particle final lateral positions with respect to their releasing positions for the four different red blood cell initial configurations. {\bf c.} Module of the average maximum displacement exerted by the particles (both, rigid and soft) with respect to their releasing positions. {\bf d.} Mean particle lateral velocity classified with respect to particle releasing positions. {\bf e.} Number of particles in time populating the three different flow regions (bulk, intermediate and cell-free layer) averaged over the four red blood cell configurations. {\bf f.} Average number of firmly adhering, interacting with the vasculature (rolling or adhering) and marginating particles over the four different red blood cell configurations.}
\label{particleChar}
\end{figure}

Here, a thorough characterization of particle margination and adhesion abilities is given in terms of local and averaged quantities within the four RBC configurations. Firstly, the final lateral positions of all particles is drawn against their releasing positions in \textbf{Figure.\ref{particleChar}.a} and \textbf{.\ref{particleChar}.b}, for rigid and soft particles, respectively. If no cells traversed the capillary, due to the small Reynolds number and the small particle dimensions, almost null lateral displacements would be observed and the distribution of $y_{fin}/H$ would overlap the bisector of the plotted plane. On the contrary, due to the presence of cells, these points form a cloud around the bisector. In particular, regardless from the particles' stiffness, small displacements are measured when transporting particles in the 1$^{st}$ RBC configuration (black dots in \textbf{Figure.\ref{particleChar}.a} and \textbf{Figure.\ref{particleChar}.b}) while for the other three configurations the largest lateral displacements are observed for particles released into the bulk zone $0.3\le y_{0}/H\le 0.7$ (symbols in \textbf{Figure.\ref{particleChar}.a} and \textbf{.\ref{particleChar}.b}). The maximum radial displacement observed in the computations, averaged over all transported particles in the four RBC configurations is drawn in \textbf{Figure.\ref{particleChar}.c} as a function of the flow region in which particles are initially set. Interestingly, the two particle families behave differently when released into the intermediate zone and in the cell-free layer. In fact, when released into the intermediate zone, rigid particle exert, at the most a lateral dislodging of $\frac{\langle r_{\max }\rangle }{H}=0.147\pm 0.052$, while for soft particle $\frac{\langle r_{\max}\rangle }{H}=0.115\pm 0.017$ (green bars in \textbf{Figure.\ref{particleChar}.c}). Then, rigid particles released into the cell-free layer move up to $\frac{\langle r_{\max }\rangle }{H}=0.050\pm 0.0047$ while soft particles move toward the centerline by about $\frac{\langle r_{\max }\rangle}{H}=0.095\pm 0.0056$ (orange bars in \textbf{Figure.\ref{particleChar}.c}). These small but statistically significant differences are due to the role of the stiffness, as already underlined in the previous sections; rigid particles are strongly scattered by RBCs while soft particles are dislodged away gently from the blood cell bulk. On the other side, rigid particles, if released into the cell-free layer, tend to avoid any direct bump with RBCs and to not move into any fluid regions perturbed by the presence of cells. The average lateral velocity measured in the computations for rigid and soft particles is drawn in \textbf{Figure.\ref{particleChar}.d} and confirms the trends depicted. The average number of particles populating the three discussed flow regions in time demonstrate that regardless from their rigidity the particles would tend to migrate out of the intermediate zone and marginate with about the same ratio (\textbf{Figure.\ref{particleChar}.e}). Lastly, over the total number of particles found at $\frac{t u_{\max }}{H}=10$ in the cell-free layer, the number of particles interacting with the endothelium is about 26\% for rigid constructs and 24\% for soft membranes. Only the 6\% of the rigid particles can firmly adhere the vasculature \textit{versus} the 23\% of the soft particles (\textbf{Figure.\ref{particleChar}.f}).

The depicted complex mechanics explains the rationale behind the choice of soft nanomedicines over their rigid counterpart. In this specific cases, the margination ability of particles seems not depend on their stiffness while their adhesive abilities are dramatically influenced by the rigidity. These trends are, in turn, confirmed by experimental findings of M. B. Fish and colleagues~\cite{fish2017}. They measured the accumulation of rigid and soft particles on vasculature walls for different flow rates and particle sizes concluding that soft particles are preferable for low shear rates while rigid constructs for high-speed flows. Note that, in this context, the overall accumulation of vascular targeted drug carriers depends on their margination and adhesion abilities at the same time. 

\subsection{\textsc{Adhesion mechanics of particles in the cell-free layer.}}

Particles sufficiently near endothelial walls are captured by the ligand-receptor mediated interaction presented in the \textbf{Computational Method} section. We now focus on the adhesion mechanics of representative rigid and soft particles in the cell-free layer (CFL) for three different values of the bond strength, $\sigma=$ 0.5, 1, and 2. The flow into the CFL is assimilated to a planar Couette flow at Re = 0.003 ($=\frac{H_{CFL}u_{\max }}{\nu}$), with $H_{CFL}=3 \mu m$. The computational domain is a rectangle in which the height is discretized with $H_{CFL}=200 \Delta x$ with $\Delta x = 15 nm$ and length correspond to $3 H_{CFL}$ with periodic boundary conditions along x.

\begin{figure}
\centering
\includegraphics[scale=0.3]{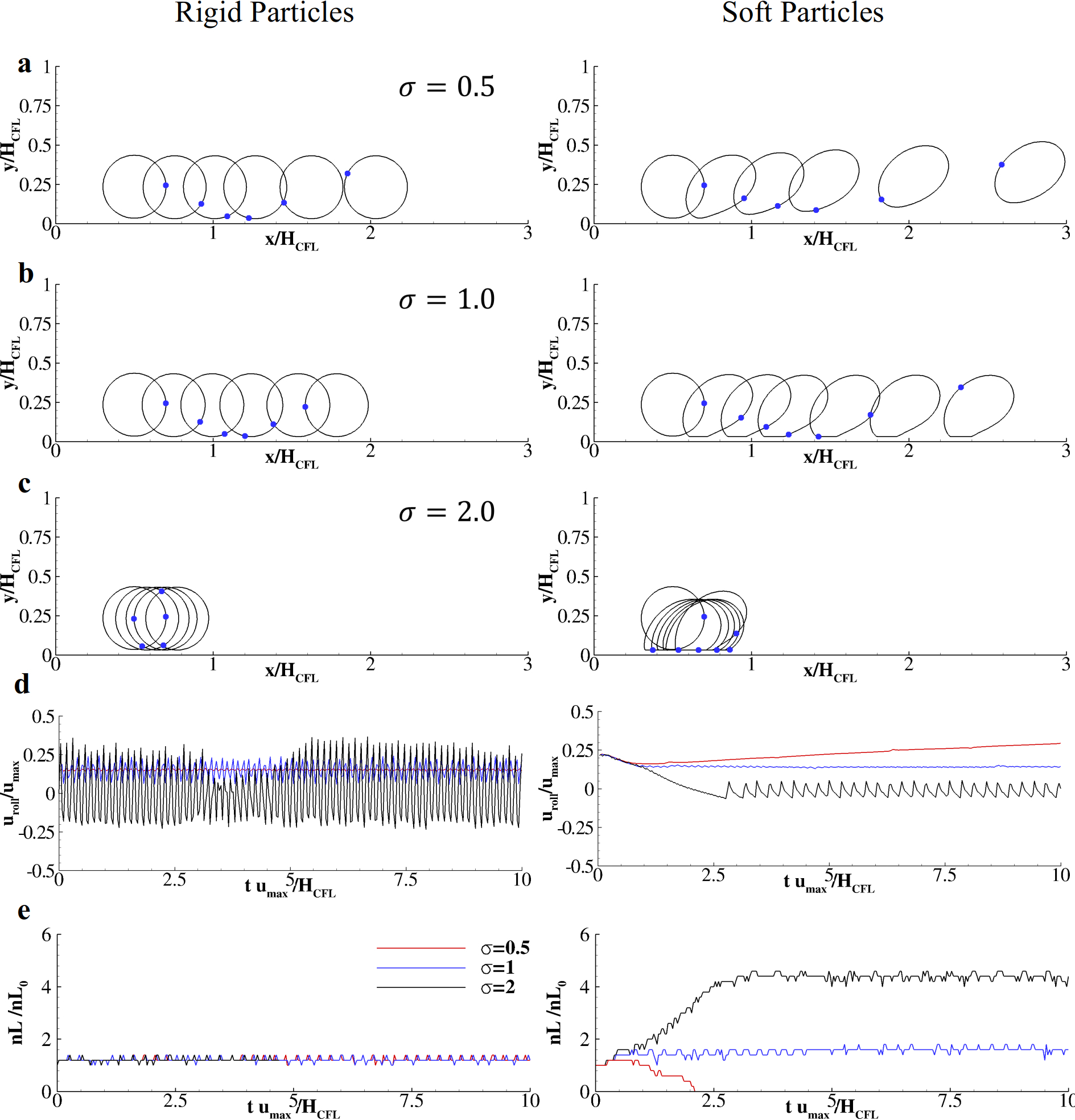}
\caption{{\bf Adhesion mechanics of rigid and soft particles navigating the cell free layer.} ({\bf a}, {\bf b}, {\bf c}) Rigid and soft particle’s configurations representing the interaction with the vasculature for three different bonds strength ($\sigma =$ 0.5, 1, and 2) taken at $\frac{t\, u_{max}}{H}=$ 0, 1, 2, 3, 4, and 5. (Blue dots are reference points for reader convenience.) {\bf d.} Rolling velocity distribution of rigid (left) and soft (right) particles with $\sigma =$ 0.5, 1.0, and 2.0. {\bf e.} Number of ligands activated during the transport of a representative rigid (left) and soft (right) particle with $\sigma =$ 0.5, 1.0, and 2.0.}
\label{adhesion}
\end{figure} 

Rigid particles preferentially roll over the endothelial wall with their motility restrained proportionally to $\sigma$ (left insets in \textbf{Figure.\ref{adhesion}.a}, \textbf{Figure.\ref{adhesion}.b}, and \textbf{Figure.\ref{adhesion}.c}). On the other hand, soft membranes may be or dislodged away from the vasculature (right inset in \textbf{Figure.\ref{adhesion}.a}); or roll on the endothelium continuously forming and destroying chemical bonds (right inset in \textbf{Figure.\ref{adhesion}.b}); or firmly adhering the vasculature deforming to maximize the contact surface (right inset in \textbf{Figure.\ref{adhesion}.c}). This dynamics is quantified \textit{via} rolling velocity measurement of such particles (see \textbf{Figure.\ref{adhesion}.d}). For $\sigma=$ 0.5 the rigid particle interacts with the wall rolling with $\frac{u_{roll}}{u_{\max }}=0.153\pm 0.02$. Then, increasing the bond strength the rolling velocity is restrained to $\frac{u_{roll}}{u_{\max }}=0.140\pm 0.09$ for $\sigma =$ 1.0 and $\frac{u_{roll}}{u_{\max }}=0.08\pm 0.25$ for $\sigma =$ 2.0.  On the other side, the soft particle with $\sigma =$ 0.5 weakly interacts with the endothelium until $\frac{t u_{\max}}{H_{CFL}}=2.3$ then detaches and moves towards higher velocity streamlines; for $\sigma =$ 1.0 the particle stably rolls with $\frac{u_{roll}}{u_{\max }}=0.142\pm 0.01$; and for $\sigma =$ 2.0 their velocity oscillate around zero, $\frac{u_{roll}}{u_{\max }}=0.0\pm 0.05$ after a short transitory period (right inset in \textbf{Figure.\ref{adhesion}.d}). The adhesion mechanics of such particles is further characterized by computing the number of closed chemical bonds in time with respect to the number of bonds originally closed in the initial configuration, $n_{L}/n_{L,0}$ (\textbf{Figure.\ref{adhesion}.f}). Indeed, rigid particles establish about the same number of bonds ($\frac{n_{L}}{n_{L,0}}=1.21$) regardless from $\sigma$. On the other hand, by varying the contact region, soft particles may vary the number of closed bonds from zero for $\sigma=$ 0.5 to $\frac{n_{L}}{n_{L,0}}=1.5$ and $\frac{n_{L}}{n_{L,0}}=4.6$ for $\sigma =$ 1.0 and 2.0, respectively, thus establishing firm adhesion with the vascular walls.

\textcolor{red}{The mechanical stiffness and the specific surface properties of such constructs are two of the four design parameters for tailoring disease-specific carriers.~\cite{decuzzi2016} It has to be noticed that, rigid and soft particles return two different adhesion dynamics. Higher mechanical stiffness enhance particles rolling, while lower values of constructs stiffness improve firm adhesion. Similar mechanisms were already observed when analyzing the crawling of large cells on a single adhesive particle by Coclite et al.~\cite{coclite20183}. In this work, the authors analyses the adhesion mechanism of particles with different size, shape and mechanical stiffness in narrow capillaries traversed by white and red blood cells suggesting soft nanomedicines as the optimal choice for the sustained release of therapeutic to the diseased tissue.}

\section*{\textsc{Conclusions}}

A combined Lattice Boltzmann - Immersed Boundary Method is employed to predict margination and adhesion mechanics of micrometric circular particles in narrow capillaries in presence of red blood cells occupying about 20\% of the total fluid volume (physiological hematocrit). These abilities are systematically assessed by transporting five independent configurations composed by four rigid and soft particles each, one per time, within four different red blood cell distributions (one single-filed and three non-filed disordered configurations). The perimeter of these particles is decorated with ligand molecules (with density $\rho_{L }= 0.5$) that may interact with receptors molecules laying on the endothelial walls. Moreover, the analysis of the adhesion mechanics of such micrometric circular constructs is carried out on a finer mesh ($\Delta x=15\, nm$) by transporting them with different ligand-receptor bond strength ($\sigma =$ 0.5, 1.0, and 2.0) in a planar linear flow thus imitating the flow into the cell-free layer.

Firstly, it is measured the formation of three different fluid regions when only blood cells are transported, namely: the bulk region, the intermediate zone and the cell-free layer. Then, the vascular journey of soft and rigid particles is analyzed in terms of their trajectories and their accumulation in these three regions, thus observing that rigid particles tend to marginate in the same fashion than their soft counterpart. Nonetheless, soft particles can adhere to the vasculature more efficiently than rigid spheres. In fact, only 6\% of the total number of rigid particles can firmly adhere the vasculature while this number grows up to 23\% when considering soft membranes. In narrow capillaries, particle stiffness does not affect the margination rate. On the contrary, particle stiffness strongly affects the ability to efficiently interact with the endothelium and firmly adhere walls.

Collectively, these data continue to demonstrate the abilities of soft nanomedicines for the precise and specific delivery of therapeutics.\textcolor{blue}{Moreover, the extension of the present analysis to a configuration including the modeling of porous tumoral tissue - in order to analyze particles extravasation process - will be the object of future investigations. On the other side, the proposed model represent a versatile tool for direct comparisons with experimental data by tuning almost all of the constitutive parameters through experimental measurements.}

\section*{\textsc{Competing Interests}}

The author declares no competing interests. 

\section*{\textsc{References}}
\bibliographystyle{unsrtnat}

\end{document}